\documentclass[preprint2]{aastex}

\usepackage{natbib}
\usepackage{graphicx}
\usepackage{amssymb}
\usepackage{epstopdf}
\usepackage[xindy, toc, hyperfirst=false, nolist, nostyles, sanitize={name=false,description=false,symbol=false}]{glossaries}
\glsdisablehyper
\usepackage{ctable, longtable}
\usepackage[hyperref,x11names, table]{xcolor}

\newcommand{\msun}{\hbox{$M_{\odot}$}}

\newcommand{\lsun}{\hbox{$L_{\odot}$}}

\newglossaryentry{vrad}{name={radial velocity~}, text={radial velocity}, symbol={\ensuremath{v_\textrm{rad}}}, description={radial velocity}, sort=vrad}
\newglossaryentry{vrot}{name={stellar rotation~}, name={stellar rotation}, symbol={\ensuremath{v_\textrm{rot}}}, description={radial velocity}, sort=vrot}
\newcommand{\vrot}{\glssymbol*{vrot}}
\newcommand{\vrad}{\glssymbol*{vrad}}

\newcommand{\rasc}[4]{\hbox{\ensuremath{\alpha=#1^h#2^m#3\overset{s}{.}#4}}}
\newcommand{\decl}[3]{\hbox{\ensuremath{\delta=#1^{\circ}#2\arcmin#3\arcsec}}}
\newcommand{\kpc}{kpc}

\newcommand{\kms}{\ensuremath{\textrm{km}~\textrm{s}^{-1}}}

\newcommand{\xray}{X-ray}

\newcommand{\snratio}{S/N ratio}

\newglossaryentry{angstrom}{name=\AA, description={unit of length $10^{-10}$\,m}, sort=angstrom}
\newglossaryentry{nir}{name=NIR,description={near infrared},first = {near infrared (NIR)}}
\newglossaryentry{psf}{name=PSF,description={Point Spread Function},first = {PSF}}
\newglossaryentry{fwhm}{name=FWHM,description={Full Width Half Maximum},first = {FWHM}}
\newglossaryentry{rms}{name=RMS,description={Root Mean Square},first = {RMS}}
\newglossaryentry{uv}{name=UV,description={ultra violet},first = {ultra violet (UV)}}
\newglossaryentry{halpha}{name=\ensuremath{\textrm{H}\alpha}, description={First line of the Balmer series at 6563\,\AA}, sort=halpha}
\newglossaryentry{mgb}{name={Mg \textsc{i} b}, description={Triplet at 5167\,\AA, 5173\,\AA and 5184\,\AA}}
\newglossaryentry{sobolevapprox}{name={Sobolev approximation}, description={Lines are approximation with an infinitley thin interaction region \citep[e.g. no broadening][]{1960mes..book.....S}}, first={Sobolev approximation }}
\newglossaryentry{radeq}{name={radiative equilibrium}, description={The net flux of energy between matter and radiation field is zero}}
\newglossaryentry{nebularapprox}{name={nebular approximation}, description={Assumes that the plasma condition are controlled by a central radiation source. The radiation field decreases with the distance to the source by geometrical dilution. See \citet{1978stat.book.....M} for details}}
\newglossaryentry{modnebularapprox}{name={modified nebular approximation}, description={In contrast to \gls{nebularapprox} where only geometrical dilution is taken into account, the modified nebular approximation also takes dilution by other radiative processes into account }, first={modified nebular approximation}, parent=nebularapprox}
\newglossaryentry{thompsonscat}{name={Thomson scattering}, description={Scattering of photons on low energy electrons}}
\newglossaryentry{lte}{name={LTE}, description={Local Thermodynamic Equilibrium}, first={local thermodynamic equilibrium (LTE)}}
\newglossaryentry{lsr}{name={LSR}, description={Local Standard of Rest}, first={\textit{local standard of rest} (LSR)}}
\newglossaryentry{mc}{name={MC}, description={Monte Carlo}, first={\textit{Monte Carlo} (MC)}}

\newglossaryentry{sfit}{name=SFIT, text=\textsc{sfit}, description={spectral fitting program for hot stars \citep{2001A&A...376..497J}}, first={\textsc{sfit} \citep{2001A&A...376..497J}}}
\newglossaryentry{iraf}{name=IRAF, text=\textsc{iraf}, description={Image Reduction and Analysis Facility maintained by NOAO}, first={\textsc{iraf}\protect\footnote{IRAF: the Image Reduction and Analysis Facility is distributed by the National Optical Astronomy Observatory, which is operated by the Association of Universities for Research in Astronomy (AURA) under cooperative agreement with the National Science Foundation (NSF).}}}
\newglossaryentry{pyraf}{name=PyRAF, text=\textsc{PyRAF}, description={Python wrap of \gls{iraf} maintained by STSCI}, first=\textsc{PyRAF} \protect\footnote{PyRAF is a product of the Space Telescope Science Institute, which is operated by AURA for NASA.}}
\newglossaryentry{scipy}{name=SciPy, text=\textsc{Scipy}, description={Scientific Python \cite{Jones:2001fk}}}
\newglossaryentry{moog}{name=MOOG,text={\textsc{moog}}, description={spectral synthesis software \citep{1973ApJ...184..839S}}, first={\textsc{Moog} \citep{1973ApJ...184..839S}}}
\newglossaryentry{atlas9}{name=ATLAS9,description={grid of stellar atmospheres \citep{2004astro.ph..5087C}}, first={ATLAS9 \citep{2004astro.ph..5087C}}}
\newglossaryentry{vald}{name=VALD,description={Vienna Atomic Line Database \citep{2000BaltA...9..590K}}, first={Vienna Atomic Line Database \citep[VALD;][]{2000BaltA...9..590K}}}
\newglossaryentry{sextractor}{name=SExtractor, text=\textsc{SExtractor}, description={Source Extractor photometry program \citep{1996A&AS..117..393B}}, first={\textsc{SExtractor} \citep{1996A&AS..117..393B}}}
\newglossaryentry{idl}{name=IDL,text={\textsc{idl}}, description={Interactive Data Language}}
\newglossaryentry{makee}{name=MAKEE,text=\textsc{makee}, description={MAuna Kea Echelle Extraction by Tom Barlow available}}
\newglossaryentry{minuit}{name=MINUIT,text={\textsc{minuit}}, description={collection of numerical optimization tools \citep{James:1975dr}}}
\newglossaryentry{migrad}{name=MIGRAD,text={\textsc{migrad}}, description={numerical gradient optimization tools - part of \gls{minuit}}}

\newglossaryentry{2mass}{name=2MASS,description={Two Micron All Sky Survey \citep{2006AJ....131.1163S}}}
\newcommand{\twomass}{\gls{2mass}}
\newglossaryentry{scp}{name=SCP,description={Supernova Cosmology Project, led by Saul Perlmutter}, first={Supernova Cosmology Project (SCP)}}
\newglossaryentry{hzsns}{name=HZSNS,description={High Z Supernova Search, led by Brian Schmidt}, first={High Z Supernova Search (HZSNS)}}
\newglossaryentry{vlt}{name=VLT,description={Very Large Telescope located on Cerro Paranal (Chile)}, first={Very Large Telescope (VLT)}}
\newglossaryentry{flames}{name=FLAMES,description={Multi-object, intermediate and high resolution spectrograph mounted on the  \gls{vlt}}}
\newglossaryentry{hires}{name=HIRES, description={High Resolution Echelle Spectrometer mounted on the Keck Telescope}, first={High Resolution Echelle Spectrometer \citep[HIRES;][]{1994SPIE.2198..362V}}}
\newglossaryentry{lris}{name=LRIS,description={Low Resolution Imaging Spectrometer mounted on the Keck Telescope}, first={Low-Resolution Imaging Spectrometer \citep[LRIS;][]{Oke95}}}

\newglossaryentry{essence}{name=ESSENCE,description={The `Equation of State: SupErNovae trace Cosmic Expansion' project \citep[ESSENCE;][]{2002AAS...201.7809G}}, first={`The Equation of State: SupErNovae trace Cosmic Expansion' \citep[ESSENCE;][]{2002AAS...201.7809G}}}
\newglossaryentry{ifu}{name=IFU,description={Optical instrument combining spectrographic and imaging capabilities, used to obtain spatially resolved spectra}, first={Integral Field Unit (IFU)}, firstplural={Integral Field Units (IFUs)}} 

\newglossaryentry{besancon}{name=Besan\c{c}on Model, description={Model of stellar population synthesis of the Galaxy, including kinematics.}}

\newglossaryentry{int}{name=INT,description={Isaac Newton 2.5\,m Telescope}, first={Isaac Newton 2.5\,m Telescope (INT)}}
\newglossaryentry{iau}{name=IAU,description={International Astronomical Union}, first={IAU}}
\newglossaryentry{chandra}{name=Chandra,description={Chandra \xray\ Observatory (space-based)}}
\newglossaryentry{hst}{name=HST,description={Hubble Space Telescope}}
\newglossaryentry{wfpc2}{name=WFPC2,description={Wide-Field Planetary Camera 2 mounted on the \gls{hst}}, first={Wide-Field Planetary Camera 2 (WFPC2)}}
\newglossaryentry{acs}{name=ACS,description={Advanced Camera for Surveys mounted on the \gls{hst}}, first={Advanced Camera for Surveys (ACS)}}
\newglossaryentry{snls}{name=SNLS,description={Supernova Legacy Survey \citep{2003AAS...203.8209P}}, first={Supernova Legacy Survey \citep[SNLS;][]{2003AAS...203.8209P}}}
\newglossaryentry{dass}{name=DASS, description={Digitized Astronomy Supernova Survey \citep{1975PASP...87..565C}}, first={Digitized Astronomy Supernova Survey \citep[DASS;][]{1975PASP...87..565C}}}
\newglossaryentry{bait}{name=BAIT, description={Berkley Automatic Imaging Telescope \citep{1993PASP..105.1164R}}, first={Berkley Automatic Imaging Telescope \citep[BAIT;][]{1993PASP..105.1164R}}}
\newglossaryentry{kait}{name=KAIT, description={Katzman Automatic Imaging Telescope \citep{2001ASPC..246..121F}}, first={Katzman Automatic Imaging Telescope \citep[KAIT;][]{2001ASPC..246..121F}}}
\newglossaryentry{loss}{name=LOSS, description={Lick Observatory Supernova Search  \citep{2000AIPC..522..103L}}, first={Lick Observatory Supernova Search \citep[LOSS;][]{2000AIPC..522..103L}}}
\newglossaryentry{ctss}{name=CTSS,description={Cal\'{a}n/Tololo Supernova Survey \citep{1993AJ....106.2392H}}, first={Cal\'{a}n/Tololo supernova survey \citep[CTSS;][]{1993AJ....106.2392H}}}
\newglossaryentry{ctio}{name= CTIO, description={Cerro Tololo Inter-American Observatory}, first={Cerro Tololo Inter-American Observatory (CTIO)}}
\newglossaryentry{ptf}{name=PTF, description={Palomar Transient Factory \citep{2009PASP..121.1334R}}, first={Palomar Transient Factory \cite[PTF;][]{2009PASP..121.1334R}}}
\newglossaryentry{batse}{name=BATSE, description={Burst and Transient Source Experiment mounted on the Compton Gamma Ray Observatory}, first={Burst and Transient Source Experiment (BATSE)}}
\newglossaryentry{bepposax}{name=BeppoSAX, description={\xray\ satellite named in honor of Giuseppe "Beppo" Occhialini}}
\newglossaryentry{rosat}{name=ROSAT, description={short for R\"{o}ntgensatellit}, first={ROSAT}}
\newglossaryentry{hete2}{name=HETE2, description={High Energy Transient Explorer}, first={High Energy Transient Explorer (HETE)}}
\newglossaryentry{gnirs}{name=GNIRS, description={Gemini Near InfraRed Spectrograph mounted on the Gemini North Telescope}}
\newglossaryentry{wifes}{name=WiFeS, description={Wide Field Spectrograph - \gls{ifu} mounted on the 2.3\,m telescope at Siding Spring Observatory}}
\newglossaryentry{swift}{name=Swift, description={Swift Gamma-Ray Burst Mission}}
\newglossaryentry{vla}{name=VLA, description={Very Large Array radio telescope located in North America}, first={Very Large Array (VLA)}}
\newglossaryentry{evla}{name=EVLA, description={Extended Very Large Array radio telescope located in North America}, first={Extended Very Large Array (EVLA)}}
\newglossaryentry{sdss}{name=SDSS, description={Sloan Digital Sky Survey}}
\newglossaryentry{dss}{name=DSS, description={Digitized Sky Survey}}
\newglossaryentry{skymapper}{name=SkyMapper, description={SkyMapper telescope \citep{2007PASA...24....1K}}, first={SkyMapper \citep{2007PASA...24....1K}}}
\newglossaryentry{panstarrs}{name=PanSTARRS, description={Panoramic Survey Telescope \& Rapid Response System \citep{2004SPIE.5489...11K}}, first={Panoramic Survey Telescope \& Rapid Response System \citep[PanSTARRS;][]{2004SPIE.5489...11K}}}
\newglossaryentry{lsst}{name=LSST, description={Large Synoptic Survey Telescope}, first={Large Synoptic Survey Telescope \citep[LSST;][]{2006AAS...209.8604P}}}
\newglossaryentry{ppmxl}{name=PPMXL, description={PPMXL Catalog of Positions and Proper Motions on the ICRS \citep{2010AJ....139.2440R}}}
\newglossaryentry{gaia}{name=GAIA, description={Global Astrometric Interferometer for Astrophysics \citep{2001A&A...369..339P}}, first={Global Astrometric Interferometer for Astrophysics \citep[GAIA;][]{2001A&A...369..339P}}}
\newglossaryentry{ligo}{name=LIGO, description={Laser Interferometer Gravitational Wave Observatory}, first={Laser Interferometer Gravitational Wave Observatory \citep[LIGO;][]{1992Sci...256..325A}}}
\newglossaryentry{aligo}{name=Advanced LIGO, description={Advanced LIGO}, sort=ligo2}
\newglossaryentry{lisa}{name=LISA, description={Laser Interferometer Space Antenna \citep{1994ESAJ...18..219J}}, first={Laser Interferometer Space Antenna \citep[LISA;][]{1994ESAJ...18..219J}}}
\newglossaryentry{ccd}{name=CCD,description={Charged Coupled Device}, first={charged coupled device (CCD)}, firstplural={charged coupled devices (CCDs)}}

\newcommand{\sn}[2]{SN~#1#2}

\newglossaryentry{sn}{name=Supernova, text={SN}, plural={SNe}, description={exploding star}, nonumberlist=true}
\newglossaryentry{snia}{name=Type~Ia (SN~Ia), text={SN~Ia}, description={Thermonuclear explosion of a white dwarf - spectra show no hydrogen but a strong silicon line},first={Type~Ia supernova (SN~Ia)}, firstplural={Type Ia supernovae (SNe~Ia)}, plural={SNe~Ia}, parent=sn, nonumberlist=true}
\newcommand{\sneia}{\glspl*{snia}}
\newcommand{\snia}{\gls*{snia}}

\newglossaryentry{branchnormal}{name={branch-normal}, text=\textit{Branch-normal}, description={Large homogeneous class of Type Ia Supernovae, defined in \citet{1993AJ....106.2383B}}, first={\textit{Branch-normal} SNe Ia \citep{1993AJ....106.2383B}}, parent=snia} 
\newglossaryentry{91t}{name={91T-like}, description={Luminous class of Type Ia supernovae similar to \sn{1991}{T} \citep{1992AJ....103.1632P}} , first={91T-like}, parent=snia} 
\newglossaryentry{91bg}{name={91bg-like}, description={Faint class of Type Ia supernovae similar to \sn{1991}{bg} \citep{1992AJ....104.1543F}}, first={91bg-like}, parent=snia} 
\newglossaryentry{02cx}{name={02cx-like}, description={Peculiar class of Type Ia supernovae similar to \sn{2002}{cx} \citep{2003PASP..115..453L}}, first={02cx-like \sneia\ \citep{2003PASP..115..453L}}, parent=snia} 

\newglossaryentry{snibc}{name=Type~Ib/c, text={SN~Ib/c}, description={Collapse of the core of a massive star -  spectrum shows no hydrogen and no silicon line},first={Type~Ib/c supernova (SN~Ib/c)}, firstplural={Type~Ib/c supernovae (SNe~Ib/c)}, plural={SNe~Ib/c}, parent=sn}

\newglossaryentry{snib}{name=Type~Ib, text={SN~Ib}, description={Spectrum shows no hydrogen and no silicon, but helium line},first={Type Ib supernova (SN~Ib)}, firstplural={Type~Ib supernovae (SNe~Ib)}, plural={SNe~Ib}, parent=snibc}

\newglossaryentry{snic}{name=Type~Ic, text={SN~Ic}, description={Spectrum shows no hydrogen, no silicon and no helium line},first={Type~Ic supernova (SN~Ic)}, firstplural={Type~Ic supernovae (SNe~Ic)}, plural={SNe~Ic}, parent=snibc}

\newglossaryentry{snii}{name=Type~II, text={SN~II}, description={Collapse of the core of a massive star - spectrum shows strong hydrogen line},first={Type~II supernova (SN~II)}, firstplural={Type~II supernovae (SNe~II)}, plural={SNe~II}, parent=sn}

\newglossaryentry{sniib}{name=Type~IIb, text={SN~IIb}, description={Spectrum shows hydrogen and helium lines},first={Type~IIb supernova (SN~IIb)}, firstplural={Type~IIb supernovae (SNe~IIb)}, plural={SNe~IIb}, see=snib, parent=snii}

\newglossaryentry{sniip}{name=Type~II~Plateau (Type IIP), text={SN~IIP}, description={Lightcurve shows plateau},first={Type~IIP supernova (SN~IIP)}, firstplural={Type~II Plateau supernovae \citep[SNe~IIP;][]{1979A&A....72..287B}}, plural={SNe~IIP}, parent=snii}

\newglossaryentry{sniil}{name=SN~II~Linear, text={SN~IIL}, description={Lightcurve shows no plateau, but linear decline},first={Type~IIL supernova (SN~IIL)}, firstplural={Type~II~Linear supernovae \citep[SNe~IIL;][]{1990MNRAS.244..269S}}, plural={SNe~IIL}, parent=snii}

\newglossaryentry{sniin}{name=Type II narrow-lined (Type IIn), description={Spectrum shows narrow lines},first={Type~II~narrow-lined supernova (SN IIn)}, firstplural={Type~IIn supernovae (SNe~IIn)}, plural={SNe~IIn}, parent=snii}

\newglossaryentry{snr}{name=Remnant (SNR), text=SNR, description={Remnant left visible post-explosion}, first={supernova remnant (SNR)}, firstplural={supernova remnants (SNRs)}, parent=sn}

\newglossaryentry{dtd}{name=DTD,description={delay time distribution - expected supernova rate over time after a brief outburst of starformation},first={delay time distribution (DTD)}, firstplural={delay time distributions (DTDs)}, plural=DTDs}

\newglossaryentry{hvg}{name=HVG,description={high velocity gradient - Type Ia supernovae with a fast evolution of photospheric velocity},first={high velocity group (HVG)}, firstplural={high velocity groups (HVGs)}, plural=HVGs, parent=snia}

\newglossaryentry{lvg}{name=LVG,description={low velocity gradient - Type Ia supernovae with a slow evolution of photospheric velocity},first={low velocity group (LVG)}, firstplural={low velocity groups (LVGs)}, plural=LVGs, parent=snia}

\newglossaryentry{wd}{name=White Dwarf (WD), text=WD, description={White Dwarf - extremely dense stellar remnant}, first={white dwarf (WD)}}
\newglossaryentry{onemgwd}{name= Oxygen/Neon (ONe), text={ONe-WD},description={Oxygen/Neon White Dwarf}, first={oxygen/neon White Dwarf (ONe-WD)}, parent=wd}
\newglossaryentry{cowd}{name=Carbon/Oxygen (CO), text={CO-WD}, description={Carbon/Oxygen White Dwarf}, first={carbon/oxygen White Dwarf (CO-WD)}, firstplural = {carbon/oxygen White Dwarfs (CO-WDs)}, parent=wd}

\newglossaryentry{sds}{name=SD-Scenario,description={single degenerate scenario (single white dwarf accreting from non-degenerate companion)}, first={single degenerate scenario (SD-scenario)}}

\newglossaryentry{dds}{name=DD-Scenario, description={double degenerate scenario (merging of two white dwarfs)}, first={double degenerate scenario (DD-scenario)}}

\newglossaryentry{sss}{name=SSS, text={supersoft \xray\ source}, description={supersoft \xray\ source - believed to be emitted by nuclear fusion on a white dwarf's surface}}

\newglossaryentry{amcvn}{name=AM CVn, description={AM Canum Venaticorum star (white dwarf accreting hydrogen poor matter from a companion star; see \cite{2005ASPC..330...27N})}}

\newglossaryentry{rlof}{name=RLOF, description={Roche Lobe Overflow (see \citet{1971ARA&A...9..183P} for a more detailed description)}, first={Roche Lobe Overflow (RLOF)}}

\newglossaryentry{mchan}{name={Chandrasekhar Mass~}, text={Chandrasekhar~mass}, symbol={\ensuremath{M_\textrm{Chan}}}, plural={Chandrasekhar~Masses}, description={Mass when the core of a star collapses due to insufficient degeneracy pressure - for a white dwarf $\approx1.38\,M_\odot$ see \citet{1931ApJ....74...81C}}, first={Chandrasekhar~mass \citep[$M_\textrm{Chan}=1.38\,M_\odot$;][]{1931ApJ....74...81C}}, sort=mchan}

\newglossaryentry{w7}{name={W7 model},description={W7 model \citep{1984ApJ...286..644N}},first = {W7 model \citep{1984ApJ...286..644N}}}

\newcommand{\smstar}{\object[NAME SM STAR]{SM-Star}}
\newcommand{\candstar}[1]{SN1006-#1}

\newglossaryentry{rp04}{name=RP04, text={RP04}, description={short for \citet{2004Natur.431.1069R}}, first={\citet[][henceforth RP04]{2004Natur.431.1069R}}}

\newglossaryentry{gh09}{name=GH09, text={GH09}, description={short for \citet{2009ApJ...691....1G}}, first={\citet[][henceforth GH09]{2009ApJ...691....1G}}}

\newglossaryentry{wek09}{name=WEK09, text={WEK09}, description={short for \citet{2009ApJ...701.1665K}}, first={\citet[][henceforth WEK09]{2009ApJ...701.1665K} }}

\newglossaryentry{ew}{name=Equivalent Width, text={EW}, description={width of a rectangle that has the same area as a spectral line when taken to zero flux}, first={equivalent width (EW)}, firstplural={equivalent widths (EWs)}}
\newglossaryentry{agb}{name=AGB,description={Asymptotic Giant Branch}}
\newglossaryentry{cmb}{name=CMB,description={Cosmic Microwave Background}}
\newglossaryentry{csm}{name=CSM,description={Circumstellar Medium}, first={circumstellar medium (CSM)}}
\newglossaryentry{csi}{name=CSI,description={Circumstellar Interaction}, first={circumstellar interaction (CSI)}}
\newglossaryentry{ism}{name=ISM,description={Interstellar Medium}, first={interstellar medium (ISM)}}
\newglossaryentry{ige}{name=IGE,description={Iron Group Element}, first={iron group element (IGE)}, firstplural={iron group elements (IGEs)}}
\newglossaryentry{epm}{name=EPM,description={Expanding Photosphere Method \citep{1974ApJ...193...27K}}, first={Expanding Photosphere Method (EPM)}}
\newglossaryentry{aic}{name=AIC,description={Accretion Induced Collapse}, first={accretion induced collapse (AIC)}}
\newglossaryentry{ime}{name=IME,description={Intermediate Mass Element}, first={intermediate mass element (IME)}, firstplural={intermediate mass elements (IMEs)}}
\newglossaryentry{h0}{name=\ensuremath{H_0},description={Hubbles constant}}
\newglossaryentry{nse}{name=NSE,description={Nuclear Statistical Equilibrium}, first={nuclear statistical equilibrium (NSE)}}
\newglossaryentry{cdm}{name=CDM,description={Cold Dark Matter}}
\newglossaryentry{grb}{name=GRB,description={Gamma Ray Burst}, first={Gamma Ray Burst (GRB)}, firstplural={Gamma Ray Bursts (GRBs)}}
\newglossaryentry{donor}{name=donor,description={non-degenerate companion in the \gls{sds}}}
\newglossaryentry{mainsequence}{name=main sequence,description={main sequence star}}
\newglossaryentry{redgiant}{name=red giant,description={red giant star}}
\newglossaryentry{mlcs}{name=MLCS,description={Multicolor Light Curve Shape method \citep[MLCS;][]{1996ApJ...473...88R}}, first={Multicolor Light-Curve Shape method \citep[MLCS;][]{1996ApJ...473...88R}}}
\newglossaryentry{rsoph}{name=RS~Ophiuci ,description={white dwarf accreting from a red giant - assumed progenitor of the \gls{sds}}, sort=rsoph}
\newglossaryentry{usco}{name=U~Scorpii,description={white dwarf accreting from a main sequence star - assumed progenitor of the \gls{sds}}, sort=usco}
\newglossaryentry{rcw86}{name=RCW~86,description={supernova remnant sometimes associated with \sn{185}{}}, sort=rcw86}
\newglossaryentry{casa}{name=Cas~A,description={Cassiopeia A supernova remnant - probably a \gls{snib} event}}
\newglossaryentry{cepheid}{name=Cepheid,description={very luminous variable star with a strong luminosity period relationship}}
\newglossaryentry{urca}{name=Urca, text=\textit{Urca}, description={process predominatly contributing to cooling in stars. The \textit{Urca} process consists of alternating electron-capture and $\beta^{-}$ decay of two nuclei pairs.},sort=urca} 
\newglossaryentry{alphacen}{name=Alpha Centauri,description={one of the brightest stars in the night sky and a close binary}}
\newglossaryentry{pcygni}{name={P Cygni}, text={P Cygni},description={a hypergiant luminous blue variable with strong winds. Often referred to as a description for their line profiles showing a emission peak at the rest wavelength of the line and a blue-shifted absorption trough.}}

\newglossaryentry{teff}{name={effective temperature~}, text={effective temperature}, symbol={\ensuremath{T_\textrm{eff}}}, description={Temperature of a blackbody emitting the same total energy}, sort=teff}

\newglossaryentry{logg}{name={surface gravity~}, text={surface gravity}, symbol={\ensuremath{\textrm{log}\,g}}, description={gravity at the surface of a star}, sort=logg}
\newglossaryentry{feh}{name={metallicity~}, text={metallicity}, symbol=\textrm{[Fe/H]},description={iron abundance relative to the sun}, sort=feh}

\newglossaryentry{texp}{name={time since explosion~}, text={time since explosion}, text={time since explosion}, symbol={\ensuremath{t_{\rm exp}}},description={time since explosion (measured in days)}, sort=texp, first={time since explosion (\ensuremath{t_{\rm exp}})}}

\newglossaryentry{lmc}{name=LMC,description={Large Magellanic Cloud}, first={Large Magellanic Cloud (LMC)}, sort=lmc}
\newglossaryentry{smc}{name=SMC,description={Small Magellanic Cloud}, sort=smc}
\newglossaryentry{z}{name=\ensuremath{z},description={redshift}, sort=z}
\newcommand{\teff}{\glssymbol*{teff}}
\newcommand{\logg}{\glssymbol*{logg}}
\newcommand{\feh}{\glssymbol*{feh}}

\renewcommand{\sn}[2]{\object{SN~#1#2}}

\usepackage{amsmath}
\usepackage{natbib}

\newcommand{\plotdir}{.}

\shortauthors{W.E. Kerzendorf et al.}
\shorttitle{Hunting for the companion of SN~1006}

\begin{document}

\title{Hunting for the progenitor of SN~1006: High resolution spectroscopic search with the FLAMES instrument%
\footnote{Based on observations collected at the European Organisation for Astronomical Research in the Southern Hemisphere, Chile (ESO 083.D-0805(A))}}
\author{Wolfgang~E.~Kerzendorf\altaffilmark{1}\altaffilmark{2}, Brian~P.~Schmidt\altaffilmark{1}, John~B.~Laird\altaffilmark{3}, Philipp~Podsiadlowski\altaffilmark{4}, Michael~S.~Bessell\altaffilmark{1}} 
\email{wkerzend@mso.anu.edu.au}

\altaffiltext{1}{Research School of Astronomy and
Astrophysics, Mount Stromlo Observatory, Cotter Road, Weston Creek,
ACT 2611, Australia}

\altaffiltext{2}{Department of Astronomy and Astrophysics, University of Toronto, 50 Saint George Street, Toronto, ON M5S 3H4, Canada}
\altaffiltext{3}{Department of Physics \& Astronomy,
Bowling Green State University,
Bowling Green, OH 43403, USA}


 \altaffiltext{4}{Department of
Astrophysics, University of Oxford, Oxford, OX1 3RH, United
Kingdom} 

\begin{abstract}

Type Ia supernovae play a significant role in the evolution of the Universe and have a wide range of applications. It is widely believed that these events are the thermonuclear explosions of carbon-oxygen white dwarfs close to the Chandrasekhar mass (1.38~\msun). However, CO white dwarfs are born with masses much below the Chandrasekhar limit and thus require mass accretion to become Type Ia supernovae. There are two main scenarios for accretion. First, the merger of two white dwarfs and, second, a stable mass accretion from a companion star. According to predictions, this companion star (also referred to as donor star) survives the explosion and thus should be visible in the center of Type Ia remnants. In this paper we scrutinize the central stars (79 in total) of the SN~1006 remnant to search for the surviving donor star as predicted by this scenario. We find no star consistent with the traditional accretion scenario in SN1006.
\end{abstract}

\maketitle
\section{Introduction}

\sneia have applications in a wide range of astronomical fields. Their iconic use as cosmological
distance probes which enabled the discovery of the accelerated expansion of the universe is augmented by their role as major drivers of chemical evolution
in the universe. They are also physically interesting endpoints of
stellar evolution. It is therefore an embarrassment that the progenitors of these explosion are as yet unknown. 

The spectra and light curves of \sneia\ suggest that the explosion is powered by burning of degenerate carbon-rich matter, suggesting carbon/oxygen white dwarfs as progenitors of SN~Ia. These objects provide carbon-rich degenerate matter and self ignite as they approach the Chandrasekhar mass threshold (1.38~\msun). Most white dwarfs, however, are born with 0.6~\msun, and how they reach the required threshold is as of yet unknown. Two main scenarios have been identified by the community \citep{1997thsu.conf..111I}. The first channel involves the merger of two white dwarfs (double-degenerate scenario) with a total mass above the Chandrasekhar limit and explosive nucleosynthesis of the merger product. In the second channel, mass is accreted from a non-degenerate companion (single-degenerate scenario). Close to the Chandrasekhar limit, explosive carbon burning ensues and the white dwarf undergoes a thermonuclear detonation/deflagration. This scenario offers an important calling card: the survival of the non-degenerate companion (also known as donor star). 

Our lack of understanding about the progenitor channel has significant impact on our understanding of the chemical evolution of the Universe. Thus, the community has put considerable effort into uncovering the progenitor scenario (for a review see Kerzendorf et al. 2012, subm.). A direct detection of a surviving donor star in a Galactic  Type Ia remnant would substantiate the single degenerate channel for at least one system.

The community has identified several Galactic Type Ia remnants that lend themselves to search for the surviving donor star (RCW86, \sn{1006}{}, \sn{1572}{} and \sn{1604}{}). In this paper, we have chosen SN~1006 for a spectroscopic search of the inner stars. The lack of a central neutron star, observations of several tenths of a solar mass of iron inside the remnant \citep{1997ApJ...481..838H}, the high peak luminosity and basic light curve shape \citep[visible for several years;][]{1965AJ.....70..748G} all indicate that \sn{1006}{} was a \snia. The remnant has a secure distance, measured by \citet{2003ApJ...585..324W}, who combined the proper motion and the radial velocity of the expanding shell to measure the distance to 2.2~\kpc, making \sn{1006}{} the closest of the Galactic \snia\ remnants (consistent with \sn{1006}{} being the brightest).   The geometric centre of the remnant is well determined from both \xray\ and radio observations \citep{2003ApJ...585..324W}. 

The interior of the remnant has been probed with UV background sources \citep{2005ApJ...624..189W}, which revealed the aforementioned iron core as well as a silicon-rich shell (adding to the evidence that \sn{1006}{} was a \snia\ event). In addition, the remnant has been searched previously for possible objects associated with the supernova explosion, which revealed an unusual O-star \citep[Schweizer-Middleditch Star (henceforth SM-Star);][]{1980ApJ...241.1039S} that  was suggested as a possible remnant star to \sn{1006}{}.  After successful identifications of neutron stars in both the Vela Remnant and the Crab Remnant this was thought to be the third identification of a stellar remnant in a historical supernova. Subsequent UV spectroscopic follow-up of the Schweizer-Middleditch star (\smstar) by \citet{1983ApJ...269L...5W} showed strong \ion{Fe}{2} lines with a profile broadened by a few thousand \kms and symmetrically distributed around the rest-wavelength. In addition, \citet{1983ApJ...269L...5W} identified redshifted \ion{Si}{2}, \ion{Si}{3} and \ion{Si}{4} lines. Their conclusion was that these absorption lines stem from the remnant and place the \smstar\ behind the remnant, making it unrelated to \sn{1006}{}. The \smstar\ remains an ideal object to probe the remnant and measure upper limits for interstellar extinction E(B-V) = 0.1 \citep[][]{1993ApJ...416..247W,2003ApJ...585..324W}.

 \sn{1006}{} has several properties which make it well suited for a progenitor search.  Although the remnant is the oldest among the remnants with a secure Type Ia identity, its age is still young enough that the remnant's centre is well determined, and the motion of any potential donor star is low enough that only a small area needs to be searched. Furthermore, the elapse of 1005 years is a short length of time relative to the timescales of stellar evolution for donor stars \cite[see][]{2000ApJS..128..615M} so we still expect a potential donor star to be close to the same state as directly after the supernova explosion. In addition, \sn{1006}{} has a low interstellar extinction, which eases the determination of stellar parameters. Finally, with 2.2\,\kpc\ it is the closest of the known \gls{snia} remnants and thus is suitable for a relatively deep survey. These serendipitous conditions for the \sn{1006}{} remnant led us to launch a photometric and spectroscopic campaign to search for the donor star.

In Section~\ref{sec:obs_red} we outline the observations as well as data reduction of the photometric and spectroscopic data. Section~\ref{sec:sn1006_analysis} is split into three subsections, namely radial velocity, stellar rotation and stellar parameters. We discuss our findings in Section~\ref{sec:sn1006:discussion} and conclude our work in Section~\ref{sec:sn1006:conclusion}.

\section{Observations and Data Reduction}
\label{sec:obs_red}
\subsection{Photometric Observations}
CCD images of the central 6\arcmin\ of the \sn{1006}{} remnant were obtained using the imaging camera at the Nasmyth-B focus of the ANU 2.3~m Telescope at
the Siding Spring Observatory on 11 May 2004. The camera has a 6\arcmin diameter circular field with a scale of 0.375\arcsec/pixel. We used the broad-band Bessell UBVI filters and exposed for 1860~s in U, 1490~s in B, 788~s in V and 1860~s in I. For calibration purposes we took images of the PG1633 and PG1047 standard star regions \citep{1992AJ....104..372L} in the same filters. The seeing ranged between 1\arcsec and 2\arcsec, and the conditions were photometric. 
The data were bias corrected and flatfielded (with skyflats) using \gls{pyraf} and \gls{iraf}.

For our photometric data reduction we fitted an astrometric solution using the astrometry from the \twomass\ point source catalogue \citep{2006AJ....131.1163S} to our frames. 
We used \gls{sextractor} to measure the magnitudes of the objects in the frames (using a 2\arcsec\ aperture),  corrected for atmospheric extinction and then we calibrated our photometry to the standard Johnson-Cousins UBV(Ic) system using the Stetson magnitudes \footnote{This research used the facilities of the Canadian Astronomy Data Centre operated by the National Research Council of Canada with the support of the Canadian Space Agency} of the stars in the standard fields PG1633+099 and PG1047+003 \citep{2009AJ....137.4186L}. The measured magnitudes were supplemented with near infrared magnitudes from the \twomass\ point source catalogue (see online tables).

We have also computed temperatures from photometric colours by using the polynomials given in \citet{2010A&A...512A..54C}. In the first instance, we assumed a solar \gls{feh} for all stars, but the choice of \gls{feh} only has a relatively minor influence on the temperature determination, for example, using B-V there is a change of less than 200~K for metallicities between \feh=0 and \feh=-1, while for V-K and V-I there is virtually no change. All temperatures are provided in online tables.

\subsection{Spectroscopic Observations}

For the spectroscopy survey we used the \gls{vlt} instrument \gls{flames}, which can provide high-resolution ($R = 25,000$) optical spectra over a 25\arcmin\ field of view for up to 130 objects. In this mode, the spectral coverage is limited to 200~\AA. We chose the wavelength region from 5139~\AA\ to 5356~\AA\ which contains the gravity sensitive \gls{mgb} triplet as well as many iron lines, to accurately measure metallicity. For the centre of our spectroscopic survey we chose the mean of the \xray\ and radio centre \citep[\rasc{15}{02}{22}{1}\ \decl{-42}{05}{49};][]{2003ApJ...585..324W}. We do note that the center choice is one of the most problematic choices in a progenitor search. Particularly, measurements by \citet{2005ApJ...624..189W} cast doubt on a precise determination of the centre of \sn{1006}{}. Their research suggests that the centre of the iron core is offset from the geometric centre determined by the shocked \gls{ism}. However, we argue that this does not mean that the centre of mass (where a donor star would reside) is necessarily off centre. In fact, \cite{2010ApJ...708.1703M} suggest that the iron ejecta is offset from the centre of mass, which suggests that the centre of the iron core will be different from the centre of mass. In general, explosion models are consistent with the center of mass being given by the outer shock, not the iron core. In addition, we chose a generous search radius of 120\arcsec, corresponding to the motion of a star travelling 1250~\kms at 2.2~\kpc\ over 1000 years. This choice, which is more than four times the maximum expected escape velocity of the donor from the system \citep{2008ApJ...677L.109H}, was made to accommodate any errors in the choice of the centre. Although the models predict the surviving companion to be several hundred \lsun\ \citep{2000ApJS..128..615M}, we chose a limiting magnitude of $V=17.5$ ($0.5~\lsun(V)$ at 2.2~\kpc\ including reddening of E(B-V)=0.1; see Equation \ref{eqn:lum_dist}) to accommodate a wide range of potential \gls{donor} star scenarios. 
\begin{equation}
\label{eqn:lum_dist}
L_{\odot}(V) = 10^{-0.4(V - 4.83) + 1.24 * E(B-V)} 4.8\times10^4 \frac{d}{2.2\,kpc}^2
\end{equation}

\begin{figure*}[tb!] 
   \centering
   \includegraphics[width=\textwidth, trim=2cm 0 4cm 0, clip]{\plotdir 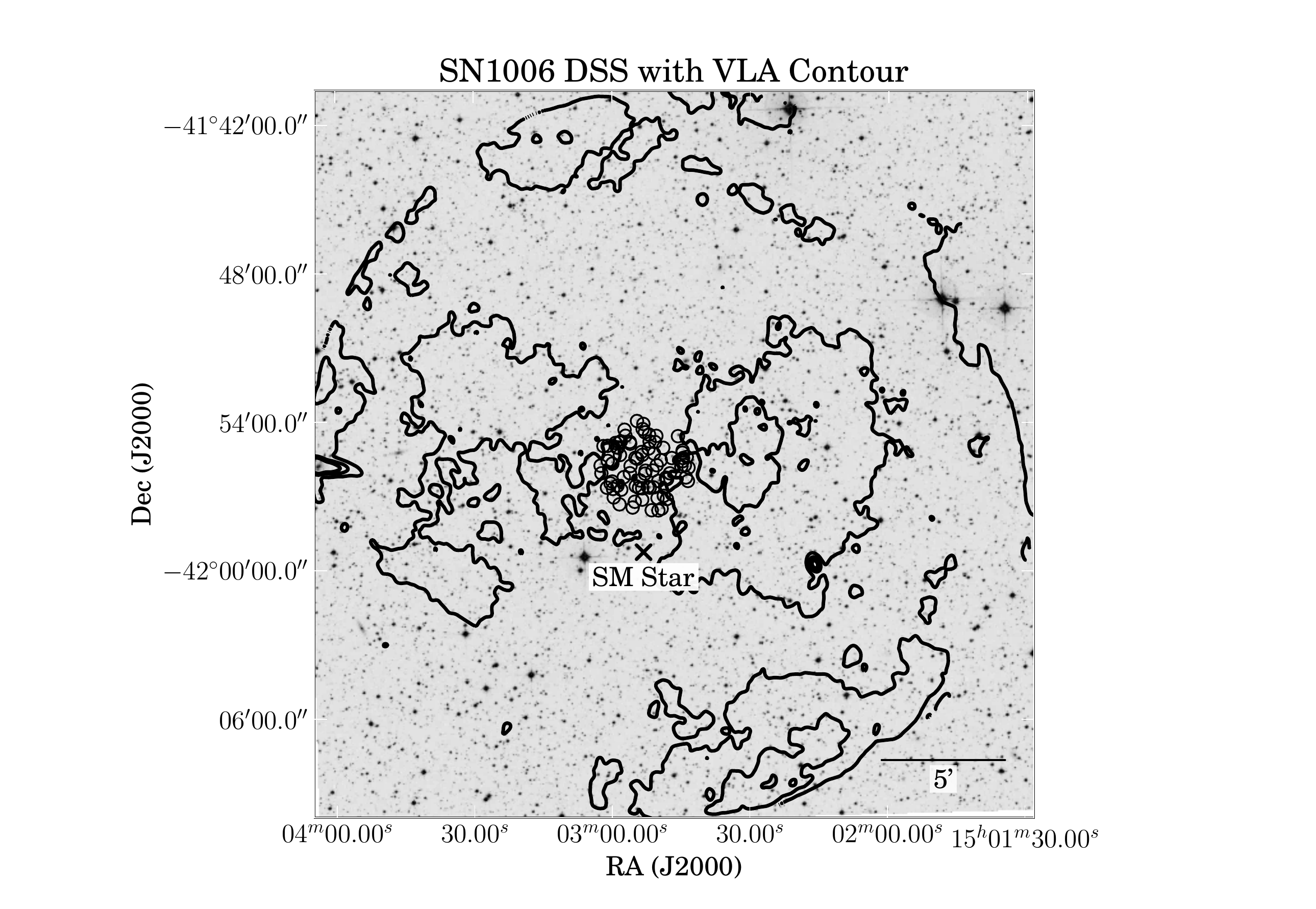} 
   \caption[Overview of candidates and remnant in SN 1006]{Optical DSS image with radio contour overlay \citep[VLA data from][]{1993AJ....106.1566M}. The black circles in the centre show the 79 program stars. Additionally we have marked the `spurious` donor the Schweizer-Middleditch star.}
   \label{fig:overview_sn1006}
\end{figure*}

\begin{figure*}[tb!] 
   \centering
   \includegraphics[width=\textwidth, trim=2cm 0 5cm 0, clip]{\plotdir 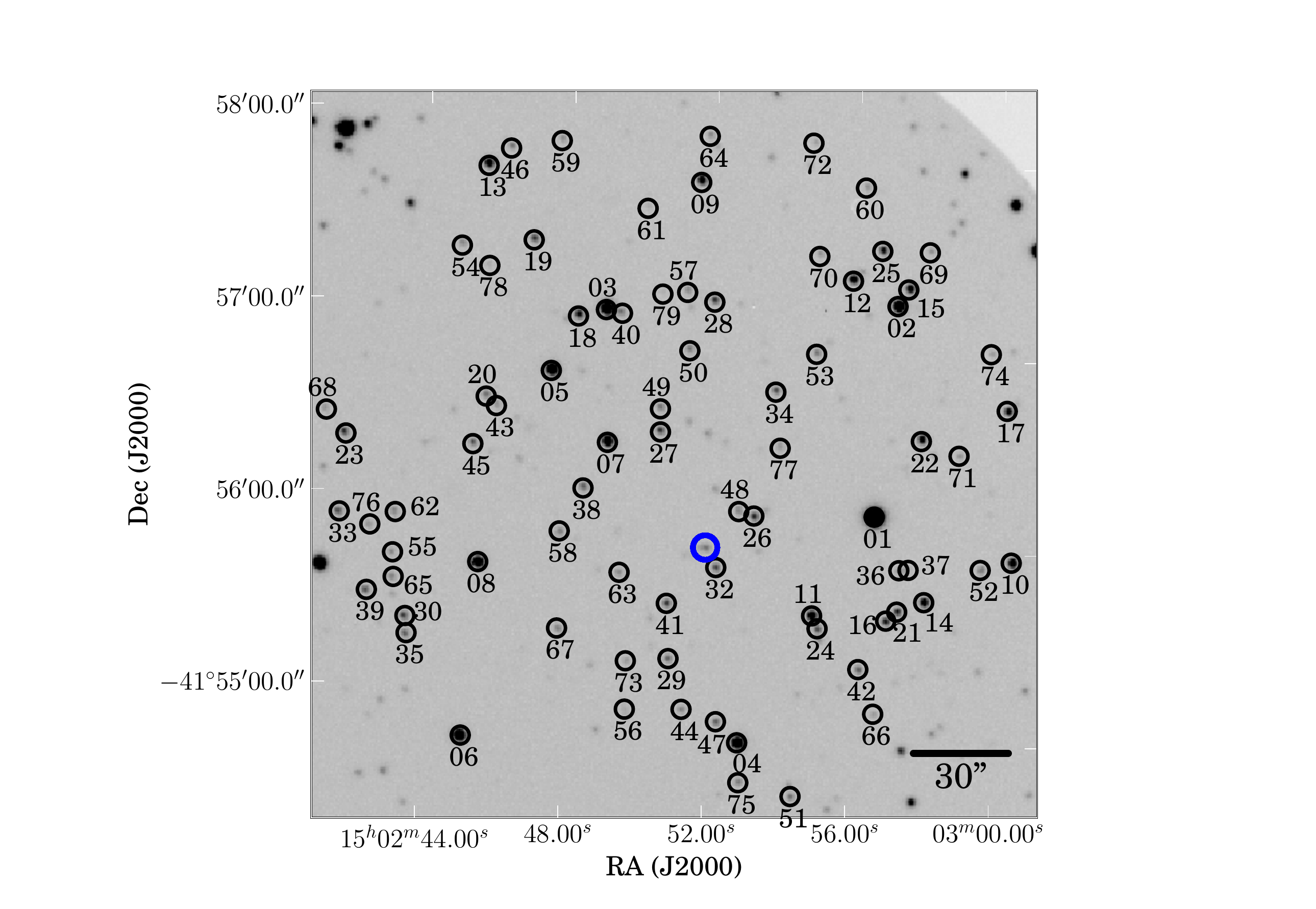} 
   \caption[Close-up of the candidates in SN 1006] {V-Band image taken by the 2.3~m Telescope. We have marked \candstar{31}, which was not observed due to broken fibres, with a blue circle. With the a brightness of $V=17.87$, \candstar{31} is fainter than our primary catalog ($V<17.5$~mag), and is the only star which lacks a spectrum to $V=19~\textrm{mag}$ in the remnant's centre.}
   \label{fig:sn1006:zoomed_overview}
\end{figure*}

An exposure time of 3.8~hours was chosen to obtain spectra with high enough quality to measure rotation and basic stellar parameters (\snratio\ $>20$). For completeness and so to not waste fibres we chose additional stars down to a magnitude limit of $V=19$ only to be used for radial velocity measurements. There are 26 stars with $V<17.5$ and 53 stars with $17.5<V<19~\textrm{mag}$ (for a total of 79 stars) for our survey (see Figure \ref{fig:overview_sn1006}). With fibre buttons not being able to be placed less than 11\arcsec\ apart, we had to split our candidates over three different setups. The first two setups were observed five times for 2775~s each. We deliberately chose bright stars for the last setup so that it only had to be observed three times for 2775~s each. In addition, we placed spare fibres on three bright stars (R$\approx 10$; \object{2MASS J15032744-4204463}, \object{2MASS J15031746-4204165}, \object{2MASS J15033195-4202356}) located close to the edge of the 25\arcmin\ field of view for calibration purposes. Additional spare fibres were placed on sky positions, which were chosen to be far from \twomass\ sources and manually inspected on DSS images to be in star free regions.
In addition to our nighttime calibration, which included simultaneous arc exposures with four fibres for each observation block, we received standard daytime calibrations. In total, 13 observation blocks with an exposure time of 2775~s each were obtained. Table \ref{tab:observations} provides the Observing ID, modified Julian date, mean seeing, mean airmass, setup name and heliocentric correction for all observations (all data are available under ESO Program ID: 083.D-0805(A)). Due to broken fibres, not all stars were observed for the expected length of time. Broken fibres caused \candstar{31} not to be observed at all in this project (see Figure \ref{fig:sn1006:zoomed_overview}), although with $V=17.87$, \candstar{31} is not part of our primary sample and thus not crucial for our search.

\ctable[
caption=FLAMES Observations of SN1006 program stars,
label=tab:observations,
pos=tb!,
star
]
{cccccc}{}{\FL
ObsID & MJD & FWHM & Airmass & Setup name & $v_\textrm{helio}$ correction\\ 
- & d & \arcsec & - & - & \kms\ML
 360737 & 54965.1 & 1.2 & 1.2 & SN1006 1 & 1.5\\
360739 & 54965.1 & 1.2 & 1.1 & SN1006 1 & 1.5\\
360740 & 54965.1 & 1.0 & 1.1 & SN1006 1 & 1.4\\
360741 & 54985.0 & 0.7 & 1.4 & SN1006 1 & -7.4\\
360742 & 54964.2 & 1.5 & 1.1 & SN1006 1 & 1.7\\
360743 & 54985.0 & 0.8 & 1.2 & SN1006 2 & -7.5\\
360745 & 54985.0 & 0.9 & 1.1 & SN1006 2 & -7.6\\
360746 & 54985.1 & 1.0 & 1.1 & SN1006 2 & -7.7\\
360747 & 54985.1 & 1.0 & 1.1 & SN1006 2 & -7.7\\
360748 & 54985.2 & 0.9 & 1.1 & SN1006 2 & -7.8\\
360749 & 54963.1 & 1.2 & 1.2 & SN1006 3 & 2.4\\
360751 & 54963.1 & 1.1 & 1.1 & SN1006 3 & 2.3\\
360752 & 54963.2 & 1.1 & 1.1 & SN1006 3 & 2.3\\
\LL}

We first applied a cosmic ray removal tool on the raw 2D frames \citep{2001PASP..113.1420V}. The data were then reduced with the ESO-CPL pipeline (version 5.2.0), using the GIRAFFE instrument recipes (version 2.8.9). The only variation that was made to the default parameters was the usage of the Horne extraction algorithm instead of the ``Optimal''-extraction algorithm. This yielded 366 individual spectra of the candidate stars and an additional 39 calibration star spectra.

\section{Analysis}
\label{sec:sn1006_analysis}
\subsection{Radial Velocity}

\begin{figure*}[tb!] 
   \centering
   \includegraphics[width=\textwidth]{\plotdir 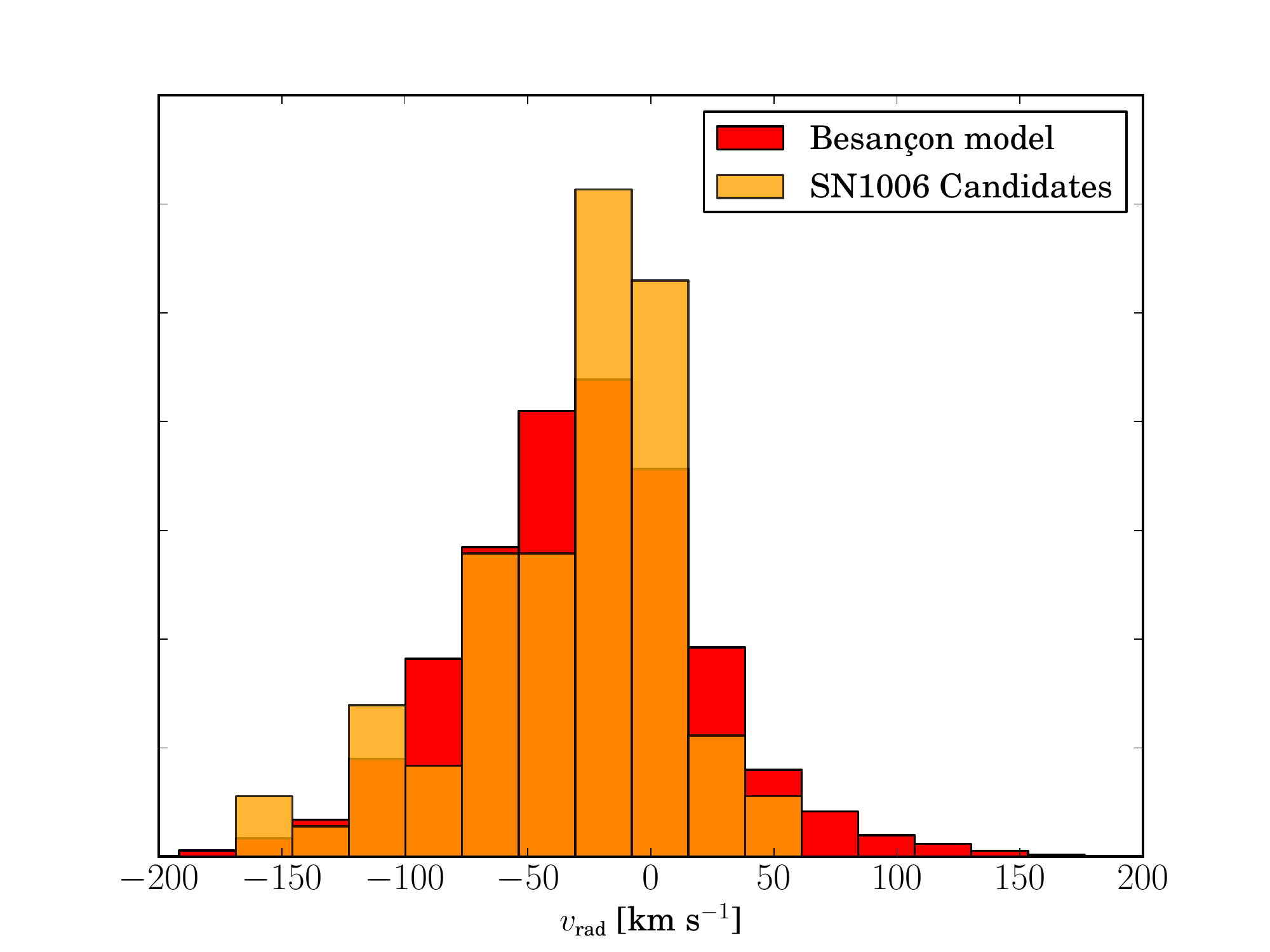} 
   \caption[Radial velocity of all candidates in SN 1006 compared with Besan\c{c}on Model]{Comparison of all candidate stars with the distribution of stars taken from the the Besan\c{c}on kinematic model. The model input parameters were a search area of 1 square degree around the centre of \sn{1006}{} and a magnitude limit of $10<\textrm{V}<17.5$}
   \label{fig:sn1006_vrad_comp}
\end{figure*}

To obtain radial velocities we employ a two-step process. We first cross correlated each spectrum with a solar spectrum from \citet{1984sfat.book.....K} using the standard  technique described in \citet{1979AJ.....84.1511T} and implemented in the \gls{pyraf} task \textsc{fxcor}. The cross-correlation was performed on each individual spectrum. In the second step, heliocentric corrections were applied and then the results were averaged for each star with a sigma clipping algorithm (for candidate stars with $V<17.5$ see Table \ref{tab:sn1006_stel_param}; the radial velocities of all stars are available online). We note that especially for faint objects we observe a second cross-correlation peak at 0~\kms and suspect that this stems from scattered moonlight. We believe that this has a negligible effect on our radial velocity measurement.
In Figure \ref{fig:sn1006_vrad_comp} we compared our radial velocity measurements with the Besan\c{c}on kinematic model of the Milky way \citep{2003A&A...409..523R}. Our selection criteria from the  Besan\c{c}on kinematic model was all stars within 1 square degree of \sn{1006}{} and a magnitude cut of $10<\textrm{V}<17.5$. We compared the resulting 10000 stars to our 78 stars in the sample in Figure \ref{fig:sn1006_vrad_comp}. 
All stars are consistent with what is expected from the Besan\c{c}on model and show no unregularities attributable to donor star candidates.

\subsection{Rotational Velocity}
\label{sec:sn1006_rotvel}

\citet{2009ApJ...701.1665K} suggest that a previously unrealized feature of donor stars is high rotation. Contrary to the predicted radial velocity signature which can be submerged in the velocity distribution of the Galaxy, the rotational velocities of stars is normally less than $10~\kms$ for late-type stars and the predicted high rotational velocities of donor stars should be clearly distinguishable. The measured rotational velocity includes a factor of $\sin{i}$. However, the expected rotational velocities are above $\vrot=50~\kms$ and thus we still expect to see modest rotation even with a high inclination angle. 

We have measured the rotational velocity of the  stars of the $V<17.5$ candidate stars using a cross-correlation technique \citep{1987AJ.....94.1066C}. First, we cross-correlated the stars with a synthetic spectrum matching closest in \teff, \logg\ and \feh\ space (and a intrinsic rotational velocity of $\vrot=2~\kms$).  If the cross-correlation peak was broader than expected from instrumental resolution ($\approx 12~\kms$), we attributed the extra broadening to rotation. This should yield the qualitatively correct result, although there maybe significant systematic errors (due to other broadening effects, however $<5~\kms$). However, our main goal is to identify rotators with $\vrot > 30~\kms$, which this technique does with a high degree of confidence.
The resulting estimates of $\vrot\sin{i}$ are given in Table \ref{tab:sn1006_stel_param}.

\subsection{Stellar Parameters}
\label{sec:sn1006_stelparam}

We obtained detailed stellar parameters for the donor candidates with $V<17.5$ by employing a grid-based technique with a three-dimensional grid in \teff, \logg\ and \feh. \gls{moog} was used to synthesise the spectral grid using the model stellar atmospheres by \citet{2003IAUS..210P.A20C}. Line wings were taken into account up to 8~\AA\ away from the line centre, which seemed to be a reasonable compromise between grid creation time and accuracy. For the atomic lines we merged values from the \gls{vald} with adjusted values (to reproduce the Arcturus and the Sun) from \cite{2008A&A...486..951G}. In addition, we used the measured molecular lines described in \citet{1995KurCD..23.....K}. The final grid extends from 3500~K to 7500~K in \gls{teff} with a step size of 250~K, from  0 to 5 in \gls{logg} with a stepsize of 0.5,  from -2.5 to 0.5 in \feh\ with a stepsize of 0.5 (with an extra set of points at 0.2). 

We used the appropriate sections from the Solar spectrum \citep{1984sfat.book.....K} and the Arcturus spectrum \citep{2000vnia.book.....H} to calibrate our spectral grid. We measured stellar parameters by first finding the best fitting grid point and then using the minimizer \gls{minuit} to find a minimum by interpolating between the gridpoints \citep{Barber96thequickhull}. For the Sun we obtain stellar parameters of \teff=5825~K, \logg=4.4 and \feh=-0.12 and for Arcturus we obtain stellar parameters of \teff=4336~K, \logg=1.9, \feh=-0.67 \citep[cf. \teff=4436~K, \logg=1.84, \feh=-0.54;][]{2007AJ....133.2464L}. We acknowledge the error in measurement, but believe our spectral grid to be accurate enough for distinguishing a potential donor candidate against an unrelated star (our requirement is to determine \logg to 1 dex and \teff to 1000~K accuracy).

To fit our observed spectra, we first fitted the continuum with Legendre polynomials with a maximum order of 3 and a sigma clipping algorithm to discard the lines. The order that gave the lowest \gls{rms} of the fit was adopted. We then combined the spectra using the previously measured \gls{vrad} and the computed heliocentric correction. In addition, we broadened the synthetic spectral grid with a rotational kernel for each star where applicable. These spectra were then fitted using the previously described algorithm, except that we added the $B-V$ photometric temperature as a prior. As the photometric temperature uses the metallicity as an input parameter we recalculated the photometric temperature prior to using the metallicity determined by the fit. This procedure was repeated until the gravity estimate converged to less than 0.1~dex. We believe our temperatures to be good to a few hundred K, and our surface gravities and metallicities have a systematic uncertainty of roughly 0.5~dex (much smaller than our required precision). 

The stellar parameters are given in Table \ref{tab:sn1006_stel_param}. The final set of stellar parameters shows a typical distribution of many dwarfs and a few giants. None of the stars seem to be unusual in any way. 
\ctable[
caption=SN 1006 candidates ($V<17.5$) stellar parameters,
label={tab:sn1006_stel_param},
width=\textwidth,
pos=tb!,
star
]
{lXXXXXXX}{}{\FL
Name & $\teff $ & $\logg$ & \feh & 
V&\vrad & \vrot \\ 
 & K & dex & dex& mag&\kms & \kms \ML
01 & 4285 & 2.0 & $-1.0$& $13.50$ & -109.1&$<10$\\
02 & 4001 & 0.8 & $-1.4$& $15.37$&56.2&$<10$\\
03 & 5446 & 4.0 & $-0.6$& $15.04$&5.9&$<10$\\
04 & 5347 & 4.0 & $-0.6$& $15.47$&-14.3&$<10$\\
05 & 5191 & 3.7 & $-0.6$& $15.50$&-1.1&$<10$\\
06 & 5874 & 4.5 & $-0.7$& $15.50$&-103.9&$<10$\\
07 & 4884 & 4.2 & $-0.8$& $15.90$&-76.3&$<10$\\
08 & 5954 & 4.2 & $-0.5$& $15.86$&-0.6&$<10$\\
09 & 4217 & 3.9 & $-2.5$& $16.58$&-47.0&$<10$\\
10 & 5662 & 4.3 & $-0.8$& $16.30$&-20.2&$10$\\
11 & 5489 & 4.1 & $-0.8$& $16.33$&-5.9&$<10$\\
12 & 5313 & 4.4 & $-0.9$& $16.39$&-59.8&$16$\\
13 & 5114 & 4.0 & $-0.7$& $16.49$&12.3&$<10$\\
14 & 5245 & 4.3 & $-0.7$& $16.56$&-17.0&$<10$\\
15 & 5503 & 4.2 & $-0.7$& $16.63$&-72.0&$<10$\\
16 & 4448 & 4.0 & $-1.8$& $17.26$&9.4&$14$\\
17 & 5515 & 4.4 & $-1.2$& $16.66$&-102.1&$<10$\\
18 & 5341 & 4.1 & $-0.9$& $16.77$&11.6&$12$\\
19 & 3846 & 4.1 & $-2.4$& $17.39$&47.8&$17$\\
21 & 4510 & 3.1 & $-1.3$& $17.36$&-18.5&$13$\\
22 & 6448 & 4.2 & $-0.4$& $16.71$&-22.9&$13$\\
23 & 4429 & 4.0 & $-1.8$& $17.39$&63.3&$14$\\
25 & 6119 & 4.9 & $-0.7$& $17.03$&40.7&$<10$\\
26 & 5619 & 4.0 & $-1.1$& $17.23$&-7.0&$<10$\\
27 & 5336 & 4.0 & $-1.3$& $17.47$&-52.0&$<10$\\
28 & 5379 & 4.3 & $-1.1$& $17.43$&-43.5&$<10$\\
\LL}

\section{Discussion}
\label{sec:sn1006:discussion}
\begin{figure*}[tb!] 
   \centering
   \includegraphics[width=\textwidth]{\plotdir 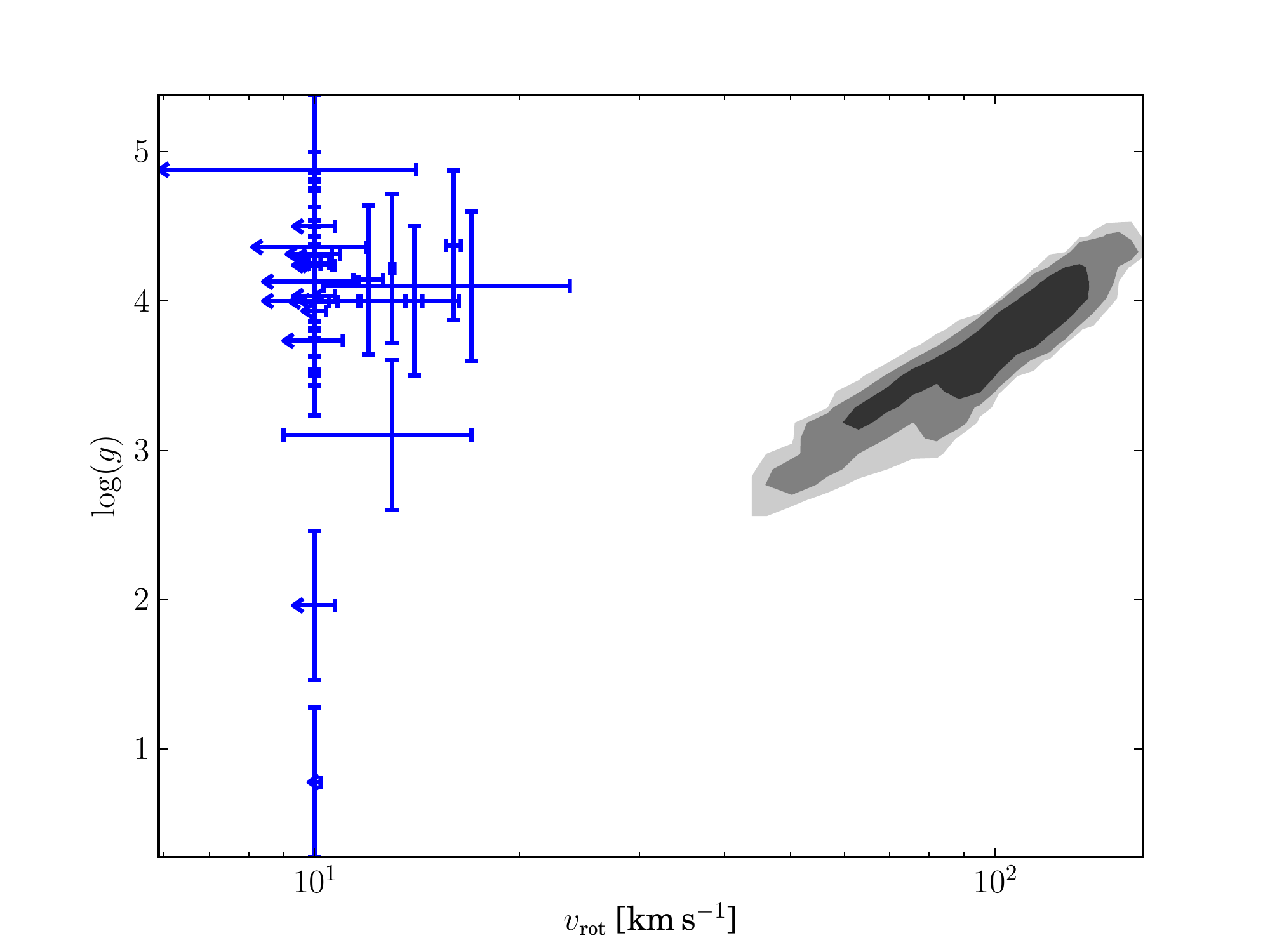} 
   \caption[Comparison of rotation and surface gravity of SN 1006 candidates]{Comparison of the evolutionary state and rotational velocity of 55000 binary synthesis \glsentryname{sds} progenitors \citep[gray shades; data from][]{2008ApJ...677L.109H} with the measured rotation from this work. Due to the resolution of the spectrograph most of these stars only have an upper limit of the rotation speed of $\vrot=10~\kms$. In addition, the measurement's are diminished by a factor $\sin{i}$ which we believe to not change the final result (expected rotation and measured rotation differ by more than a factor of 5). }
   \label{fig:han2008_vrot_compare}
\end{figure*}

In this work we have scrutinised all stars to a limit of $0.5~\lsun(V)$ at the distance of the \sn{1006}{}\ remnant. In addition, we have performed radial velocity measurements of stars down to a limit of  $\approx0.1~\lsun(V)$ at the distance of \sn{1006}{}. Although theoretical models predicted bright donor stars, we have searched down to relatively faint limits. As these predictions were only theoretical in their nature, it was important  further features that could hint at a donor star (namely rotation, radial velocity, and unusual stellar parameters). We used population synthesis models from \citet{2008ApJ...677L.109H} to judge our rotation measurement with the the rotation of donor stars post-explosion (see Figure~\ref{fig:han2008_vrot_compare}). None of the stars scrutinised in our sample show features consistent with those expected in any current donor star models or are significantly unusual.

Giant donor stars are easily ruled out because there is no star bright enough to be at the distance of the remnant. \citet{2000ApJS..128..615M} suggest that giant donors have a luminosity of $\approx 1000~\lsun$ ($V\approx9$ at the distance of the remnant) for at least 100,000 years. Furthermore, these models suggest that the giant donor is likely to have a high temperature of more than $10^4\,K$. In addition, the star should have rotation in excess of what has been measured for any of the stars in this sample. In summary, there is no viable giant star donor star among the stars located in \sn{1006}{}. 

Subgiant donors should also be very luminous post explosion \citep{2000ApJS..128..615M} with a minimum expected luminosity of $L\approx500\,\lsun$ ($V\approx9.7$ at the distance of the remnant) lasting for 1400 -- 11,000 years, although theoretical models allow a much larger variation of this class of stars \citep{2003astro.ph..3660P}. While they might have a \gls{vrad} which could be masked by the large expected dispersion in the direction of \sn{1006}{}, the expected $\vrot\approx80\,\kms$ (see Figure \ref{fig:han2008_vrot_compare}) far exceeds any star in our sample.  Therefore, we believe we can confidently rule out sub-giant donor stars in this case as well. In summary our research shows a consistent result to \sn{1572}{} - no identifiable donor star for \sn{1006}{}.

Finally, main sequence stars, according to \citet{2000ApJS..128..615M} are expected to have a similar brightness to subgiant stars, although this enhanced luminosity depends on the details of how energy is deposited from the explosion \citep[see][]{2003astro.ph..3660P}.  However, main sequence donors should have both substantial spatial motion and a very high rotation. No star, in our sample, shows any of these features and our sample's depth should cover all conceivable post-evolutionary scenarios, even for a main sequence donor star.

On caveat is that rotation can decrease due to expansion \citep{2009ApJ...701.1665K}. However, this should result in a star with a low gravity. No such star is present in SNR~1006.

\section{Conclusions}
\label{sec:sn1006:conclusion}
The observations presented here for \sn{1006}{} are in conflict with the standard \snia\ donor star scenarios, which include accretion onto a white dwarf from a main sequence, subgiant, or giant companion. 
A few non-standard scenarios survive our observational tests. These include a helium white dwarf as a donor star, which would not be detectable with our observations, although it is not entirely clear that a helium white dwarf would survive the explosion (priv. comm. R\"udiger Pakmor). In addition, \cite{2011ApJ...730L..34J} \& \cite{2011ApJ...738L...1D} suggest that the donor star might evolve and become a white dwarf before the supernova explosion. The delay in explosion is explained by the need to spin-down to reach critical density in the core of the white dwarf. This would again result in a companion, which is not detectable by our methods. Finally, another possibility is that \sneia\ (or at least \sn{1006}{}) do not have donor stars, consistent with a \gls{dds}.

\section{Acknowledgements}

B.~P.~Schmidt and W.~E.~Kerzendorf were supported by Schmidt's ARC Laureate Fellowship (FL0992131).

We would like to thank Zhanwen Han for providing population synthesis data. We would like to thank the anonymous referee for insightful and useful comments.

\bibliographystyle{hapj}
\bibliography{sn1006_flames}

\end{document}